\documentclass[prl,twocolumn,superscriptaddress,showpacs,nobalancelastpage]%
{revtex4}

\usepackage{amsmath}
\usepackage[dvips]{graphicx}

\newcommand{\D}{\ensuremath{\mathrm{d}}}
\renewcommand{\Re}{\ensuremath{\mathrm{Re}}}
\renewcommand{\Im}{\ensuremath{\mathrm{Im}}}

\newcommand{\Cee}{\ensuremath{\mathcal{C}}}

\begin{document}


\title{Lee-Yang zeros and phase transitions in nonequilibrium steady
states}
\date{18th April, 2002; revised 19th June, 2002}

\author{R.\ A.\ Blythe}
\affiliation{Department of Physics and Astronomy, University of Manchester,
Manchester, M13 9PL, UK}

\author{M.\ R.\ Evans}
\affiliation{Department of Physics and Astronomy, University of Edinburgh,
Mayfield Road, Edinburgh, EH9 3JZ, UK}

\begin{abstract}
We consider how the Lee-Yang description of phase transitions in terms
of partition function zeros applies to nonequilibrium systems.  Here
one does not have a partition function, instead we consider the zeros
of a steady-state normalization factor in the complex plane of the
transition rates.  We obtain the exact distribution of zeros in the
thermodynamic limit for a specific model, the boundary-driven
asymmetric simple exclusion process.  We show that the distributions
of zeros at the first and second order nonequilibrium phase
transitions of this model follow the patterns known in the Lee-Yang
equilibrium theory.
\end{abstract}

\pacs{05.40.-a, 05.70.Fh, 02.50.Ey}

\maketitle


With equilibrium statistical physics now firmly established as a
successful general theory for predicting the macroscopic behavior of
many-body systems, attention has more recently turned to the the wider
class of nonequilibrium systems.  Examples of nonequilibrium systems
are those relaxing to thermal equilibrium or those driven into a
steady state, far from thermal equilibrium.  A general framework for
understanding these systems is proving elusive. However, over the last
two decades the detailed study of specific models has revealed the
wide range of phenomena that may emerge. 

It is by now well known that models of driven diffusive systems can
exhibit phase transitions in their nonequilibrium steady states
\cite{SZ}.  Typically such models comprise one or more species of
particles being driven by an external field along with prescribed
interactions between the boundaries and the outside world.
Applications of these models are diverse and include the kinetics of
biopolymerization \cite{MGP}, transport across a membrane \cite{CL}
and traffic flow \cite{CSS}.  From a more fundamental viewpoint, the
interest lies in the richness of collective phenomena displayed,
including jamming \cite{OEC} and spontaneous symmetry breaking
\cite{EFGM}, even in one dimension. 

Half a century ago Lee and Yang provided a theory of equilibrium phase
transitions based around the zeros of the partition function.  In this
work we focus on nonequilibrium steady states and how the phase
transitions they admit may be placed in the context of the Lee-Yang
theory.  In the original exposition \cite{YL} the fugacity of a gas
was generalized to the complex plane and the zeros of the
grand-canonical partition function in this plane were considered.  In
particular, it was shown 
that the free energy is analytic
in any region of the
complex-fugacity plane devoid of any zeros.
Conversely, there are
nonanalyticities at points where the density of partition function
zeros accumulate in the thermodynamic limit.  These accumulation
points correspond to physically observable phase transitions if they
lie on the positive real axis.  
A similar phenomenon
occurs in the complex plane of other intensive fugacity-like
variables.  For example, Lee and Yang also showed \cite{LY} that the
zeros of the Ising model partition in the complex plane of activity
(defined as $\exp(-h/T)$ where $h$ is magnetic field and $T$
temperature) lie on the unit circle.
Similarly, partition function zeros in the complex
temperature plane also accumulate at phase transition points
\cite{Fisher,GR}.  The applicability of the Lee-Yang theory to
equilibrium transitions driven by intensive field-like variables can
be understood from the fact that, mathematically, these variables play
similar roles in the partition function. 

The difficulty in applying the Lee-Yang theory to, e.g., the steady
state of a driven diffusive system is that one does not have to hand a
partition function defined in terms of thermodynamic state variables.
However, one can always define a
quantity $Z$ as a sum of the steady state weights (unnormalized
probabilities) $f(\Cee)$:
\begin{equation}
\label{Zgen}
Z = \sum_{\Cee} f(\Cee)\;.
\end{equation}
Thus the probability of a configuration is given by $P(\Cee) =
f(\Cee)/Z$.  In this work, we treat the normalization $Z$ as a
nonequilibrium analog of the partition function. 

In order to obtain $Z$ one still has to calculate the steady state
weights. Up to a multiplicative factor they are implied by the
transition rates $W(\Cee \to \Cee^\prime)$ that define the model
through the requirement that the total inflow of probability into a
given configuration $\Cee$ must be balanced by the outflow, i.e.,
\begin{equation}
\label{steadystate}
\sum_{\Cee^\prime \ne \Cee} \left[ f(\Cee^\prime) W(\Cee^\prime \to
\Cee) - f(\Cee) W(\Cee \to \Cee^\prime) \right] = 0 \;.
\end{equation}
The solution of this set of equations for the steady-state weights
$f(\Cee)$ can, in principle, be obtained using standard methods, such
as Gaussian elimination, Cramer's rule or graphically \cite{EBt}.  If
one applies such a method, the weights can always be written as
polynomials of the elementary transition rates.  Thus, for a finite
system, $Z$ is also a polynomial of the transition rates. 

In this work we generalize the transition rates to the complex plane
and in doing so consider the zeros of the partition function.  In
analogy with the Lee-Yang picture for equilibrium systems, for a phase
transition to arise we expect the zeros to pinch the real axis at a point
which corresponds to a (real) critical transition rate. 

As an aside, we note that the general methods of solving
(\ref{steadystate}) mentioned above are rarely tractable when the
number of configurations becomes large.  One exception is when the
system satisfies detailed balance.  In this case, each term within the square
brackets in (\ref{steadystate}) is identically zero and it is easy to
construct the set of steady-state weights as Boltzmann weights
with respect to an energy function. The resulting $Z$ is then an
equilibrium partition function.
A simple example of such a
system is the Ising model evolving under Glauber dynamics
wherein studying the zeros 
of the partition function as a function of the transition rate  would
be equivalent to studying the complex-temperature
zeros. 

\begin{figure}
\includegraphics[scale=0.6]{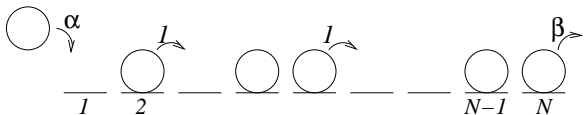}
\caption{\label{asep-fig}Dynamics of the ASEP.  The arrow
labels indicate the rates at which the corresponding transitions occur;
site labels are also indicated.}
\end{figure}
In the present work, however, we are	
interested in models which do not satisfy detailed balance and hence
for which the normalization $Z$ is not \textit{a priori} known.
This 
makes it difficult to make general statements about the significance
of the normalization $Z$.  We have argued that a nonequilibrium phase
transition will be signaled by an accumulation of the zeros of $Z$ in
the complex plane of transition rates towards the real axis.  We
demonstrate this in the case of a particular nonequilibrium model for
which an expression for $Z$ is exactly known and we can calcuate the
distribution of zeros analytically. We find that these zeros lie on
circles in the complex plane of an appropriately chosen function of a
transition rate that plays the role of a fugacity. The characteristic
properties of the distributions of partition function zeros at first
and second order equilibrium transitions \cite{YL,GR,GL} are also
exhibited in this nonequilibrium case.  In short, we show how the the
Lee-Yang theory of phase transitions generalizes to a nonequilibrium
steady state. 

Before discussing the model and our results in greater detail, we wish
to remark how our study differs from a related work due to Arndt
\cite{Arndt}.  The aim of that investigation was to locate a phase
transition induced by varying the relative numbers of different
particle species in a particle-conserving driven diffusive system.  To
this end, a fugacity-like quantity was introduced in an \textit{ad
hoc} way.  Physically, this corresponds to placing the driven
diffusive system in contact with a particle reservoir thus introducing
equilibrium fluctuations in the particle densities.  We stress that in
the present work, it is elementary transition rates, which need not
have any connection to intensive field (fugacity-like) variables, that
are generalized to the complex plane. 

We now introduce the model we consider, namely the one-dimensional
asymmetric simple exclusion process (ASEP) with open boundaries.
Since its introduction \cite{MGP,Krug} this model has received a great
deal of attention---see, e.g., \cite{Reviews,EBt} for further details
and references.  The model comprises a lattice of $N$ sites, each of
which may be occupied or empty.  In an infinitesimal time interval $\D
t$ one of the following transitions may occur: a particle may hop one
site to the right with probability $\D t$ (i.e.\ at unit rate) subject
to the receiving site being vacant; a particle may be inserted onto
the leftmost site (if vacant) with probability $\alpha \D t$ or a
particle occupying the rightmost site (if present) may be removed with
probability $\beta \D t$.  Fig.~\ref{asep-fig} illustrates the
microscopic dynamics. 

\begin{figure}
\includegraphics[scale=0.60]{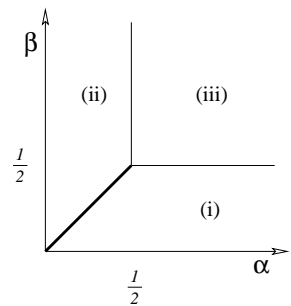}
\caption{\label{Jphase-fig}Phase diagram of the ASEP.}
\end{figure}

As outlined above, there is a current (defined as the flux of
particles between neighbouring sites per unit time) in the steady
state which in the thermodynamic limit $N \to \infty$ exhibits
nonanalyticities as the boundary rates $\alpha$ and $\beta$ are
varied.  These are associated with phase transitions, and the phase
diagram is given in Fig.~\ref{Jphase-fig}. 

Phase (i) is a high-density phase with a density profile that decays
exponentially towards the right boundary.  In this phase the current
$J=\beta(1-\beta)$.  Since the ASEP is invariant under particle-hole
exchange and the swap $\alpha \leftrightarrow \beta$, it follows that
phase (ii) is a low-density phase that has an exponential decay in the
density profile from the left boundary and a current
$J=\alpha(1-\alpha)$.  At the first order transition
line $\alpha = \beta < \frac{1}{2}$ the current
exhibits a discontinuity in its first derivative.  Here
the system exhibits a shock front separating regions
of high and low density;  this is an example of phase coexistence at a
nonequilibrium phase transition. 

In phase (iii), the current assumes a constant value $J = \frac{1}{4}$
which is the largest possible current admitted for any combination of
$\alpha$ and $\beta$.  Hence phase (iii) is called the maximal current
phase.  This phase has the interesting feature that the density
profile decays as a power-law from both boundaries and hence at the
transition from either phase (i) or (ii) to phase (iii) a correlation
length diverges.  Additionally, $J$ has a discontinuity in its second
derivative and so this is a second-order transition. 

These results are known mainly from the exact solution \cite{DEHP,SD}.
The two key results we call upon here are that the normalization for
an $N$-site system $Z_N$ is given by
\begin{equation}
\label{asepZ}
Z_N = \sum_{p=1}^{N} \frac{p(2N-1-p)!}{N!(N-p)!}  \frac{
(1/\beta)^{p+1} - (1/\alpha)^{p+1} }{ (1/\beta)-(1/\alpha) }
\end{equation}
and the current by $J = Z_{N-1}/Z_N$.  In order later to construct the
distribution of the zeros of $Z_N$ for the ASEP we also need the
large-$N$ forms of $Z_N$.  From the exact solution it can be shown
that for all $\alpha, \beta$, $Z_N \sim A J^{-N} N^{\gamma}$ where $J$
is the current and $A, \gamma$ depend only on the phase being
considered.  We see $\ln J$ is the extensive part of $-\ln Z_N$ and
thus $\ln J$ plays the role of the free energy and $J$ the fugacity
for this nonequilibrium system. 

\begin{figure}
\includegraphics[scale=0.29]{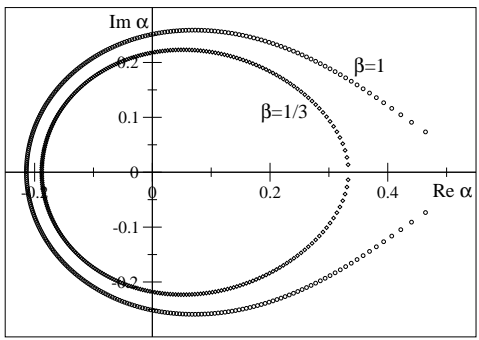}
\caption{\label{alphazeros-fig}Zeros of the normalization
(\ref{asepZ}) in the complex-$\alpha$ plane for $\beta=\frac{1}{3},1$.
In both cases the lattice size $N$ is
300.}
\end{figure}

We now consider the zeros of $Z_N$ given by (\ref{asepZ}) in the
complex-$\alpha$ plane at fixed $\beta$ \cite{DLS}.  (The symmetry in
$\alpha$ and $\beta$ of $Z_N$ implies that one would obtain the same
pattern of zeros in the complex-$\beta$ plane at fixed $\alpha$).
Using \textsc{Mathematica}, we obtained numerical estimates of the
zeros of $Z_N$ for system size (and number of zeros) $N=300$ and
$\beta=1, \frac{1}{3}$.  The results are shown in
Fig.~\ref{alphazeros-fig}. 

For the case $\beta=\frac{1}{3}$, the zeros appear to approach the
positive real $\alpha$ axis at a value $\alpha_c=\frac{1}{3}$,
coincident with the first-order phase transition point in this
regime. The plot for the case $\beta=1$ is suggestive of a slow
accumulation of the zeros to the positive real $\alpha$ axis at
$\alpha_c=\frac{1}{2}$, i.e.\ the second-order phase transition point.
It appears also in this case that the zeros approach the real axis at
an angle $\frac{\pi}{4}$ as opposed to an angle $\frac{\pi}{2}$ at the
first-order transition.  We will shortly demonstrate that the above
assertions---consistent with the patterns of zeros at first- and
second-order phase transitions known from the equilibrium Lee-Yang
theory \cite{YL,GR,GL}---are correct. 

\begin{figure}
\includegraphics[scale=0.29]{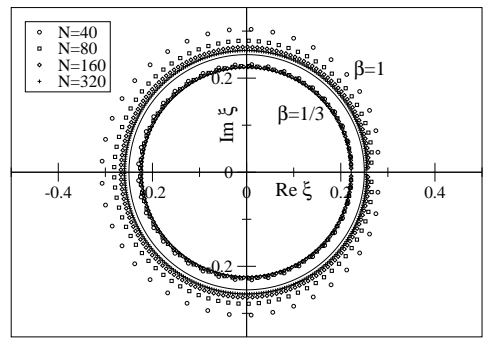}
\caption{\label{xizeros-fig} Zeros of the normalization (\ref{asepZ})
in the complex-$\xi=\alpha(1-\alpha)$ plane for $\beta=\frac{1}{3},1$
and $N=80,160,320$.}
\end{figure}

It is useful to make a change of variable from $\alpha$ to
$\xi=\alpha(1-\alpha)$ (note that in the low density phase $\xi = J$)
and we plot the numerically-obtained zeros in the complex-$\xi$ plane
in Fig.~\ref{xizeros-fig}.  For both $\beta=1, \frac{1}{3}$, it
appears that, as $N \to \infty$, the zeros become uniformly
distributed around a circle of radius $\xi_c = \alpha_c(1-\alpha_c)$
(these circles are shown as solid lines in the figure).  We shall now
show how to obtain this result analytically. 

We first note (for a proof see \cite{Derrida}) that for any polynomial
$p_N(\xi)$ with $N$ zeros, its density of zeros $\rho(x,y)$ in the
complex $\xi=x+iy$ plane is given by $\rho(x,y) = 2\pi \nabla^2 \ln
|p_N(\xi)|$ where $\nabla^2 \equiv \frac{\partial^2}{\partial x^2} +
\frac{\partial^2}{\partial y^2}$.  In the thermodynamic limit ($N \to
\infty$) we expect a continuous distribution of zeros and wish to
normalize $\rho$ so that its integral over all space is unity.  We
thus define the key quantity
\begin{equation}
\label{phidef}
\phi = \lim_{N \to \infty} \frac{\ln |Z_N|}{N}
\end{equation}
so that $\rho = 2\pi \nabla^2 \phi$.  In equilibrium statistical
physics $\phi$ would be interpreted as the extensive part of the free
energy; as noted above $\phi = -\ln |J|$ for the ASEP. 

It is useful to regard $\phi$ as an `electrostatic' potential in a
two-dimensional space.  We introduce the field $\vec{E}=\nabla\phi$
and consider the discontinuities that arise as a line of charge
(zeros) is crossed.  The appropriate forms of Gauss's and Ampere's
laws imply that the component of $\vec{E}$ parallel to a line of zeros
separating two phases must be continuous across it whereas the
perpendicular component differs by a value $2\pi g(x,y)$ where
$g(x,y)$ is the line density of charge (zeros) along the boundary. 

The current itself has only two analytic forms.  In the interior
region of Fig.~\ref{xizeros-fig} (i.e., the region that includes
$\xi=0$) $J=\alpha(1-\alpha)=\xi$ and hence $\phi=-\ln|\xi|$.  Over
the exterior region, the current is constant:
$J=\alpha_c(1-\alpha_c)=\xi_c$ and so $\phi = -\ln|\xi_c|$.  

The field that results from the potential $\phi$ is zero in the
exterior region and directed radially towards the origin in the
interior region; specifically $\vec{E} = - \vec{r}/r^2$ where $\vec{r}
= (\Re\,\xi, \Im\,\xi)$.  Now, since the component of the electric
field parallel to the boundary between the two phases varies
continuously and is zero in the exterior region, the phase boundary
must lie perpendicular to the field lines in the interior region.  In
other words, the phase boundary is an equipotential $\phi=-\ln|\xi| =
\mbox{const} \Longrightarrow |\xi| = \xi_c$ since the boundary must
pass through the transition point at $\xi=\xi_c$.  The line-density of
zeros $g(\theta)$ along the circle $\xi=\xi_c e^{i\theta}$ can be
calculated by noting that the perpendicular component of the field at
the boundary has a magnitude $1/|\xi_c|$.  The relationship given
above then implies $g(\theta) = (2\pi\xi_c)^{-1}$, i.e., a constant as
claimed. 

We now use this result to investigate the distribution of zeros near
the transition point in the complex-$\alpha$ plane.  In the $\xi$
plane, the circle of zeros passes along the line $\Re\,\xi=\xi_c$
which maps onto a curve in the complex $\alpha=u+iv$ plane that has
$(u-\frac{1}{2})^2-v^2=\frac{1}{4}-\xi_c$.  The solution of this
equation that describes a line passing through the transition point
$\alpha=\alpha_c$ is $u=\frac{1}{2}-(v^2+\frac{1}{4}-\xi_c)^{1/2}$. 

When the phase transition is first order, $\beta<\frac{1}{2}$ and
$\xi_c=\beta(1-\beta)<\frac{1}{4}$.  Then, the curve of zeros passes
smoothly through the transition point $\alpha=\beta$.  One can show
that the line density of zeros at this point in the complex-$\alpha$
plane is $(1-2\beta)/[2\pi\beta(1-\beta)]$, i.e., nonzero.  The fact
that the density of zeros is nonzero on the real axis at a first-order
phase transition is well-known in the equilibrium Lee-Yang theory
\cite{YL}. 

At the second order phase transition ($\beta \ge \frac{1}{2}$ and
$\xi_c=\frac{1}{4}$) the line of zeros is given by $u=\frac{1}{2} -
|v|$.  This means the zeros approach the transition point $\alpha=1/2$
at an angle of $\frac{\pi}{4}$ to the real axis, meeting at a right
angle.  If one defines $\ell$ as the displacement along the straight
line of zeros from the transition point, one finds that it varies with
$y=\Im\,\xi$ as $\ell = \sqrt{y}$.  This implies that the density of
zeros behaves as $g(\ell) = \frac{2}{\pi} \frac{\D y}{\D \ell} =
\frac{4\ell}{\pi}$.  In the equilibrium theory, it is well known that
the density of zeros vanishes as a power-law towards the transition
point on the real line (see, e.g., \cite{GL}), a result we have
recovered in this nonequilibrium case.  

To summarize, we have shown that the zeros for the normalization of
the ASEP accumulate at the phase transition points in the complex
plane of transition rates. As in the case of the Lee-Yang theory, a
first-order transition is manifested by a nonzero density of zeros at
the accumulation point, whereas the density of zeros decays as a
power-law towards a continuous transition point. 

A point that we have so far ignored is that
the normalization (\ref{Zgen}) is only
defined up to a multiplicative factor that depends on the method used
to solve Eqn.~(\ref{steadystate}).  This additional factor is
analogous to that which would appear in an equilibrium partition
function after a uniform shift of the energy scale.  Although this
factor could itself be a polynomial of the transition rates and so
introduce additional zeros, one would not expect these spurious zeros
to be physically relevant to the phase behavior.  For example one way
to define the normalization is to use Cramer's rule to solve
(\ref{steadystate}) (see \cite{EBt}).  Then for the ASEP the
normalization is given by $(\alpha \beta)^N$ times the expression
(\ref{asepZ}) and additional zeros are introduced only at the
origin. 

Finally we remark on the applicability of the Lee-Yang theory to other
nonequilibrium systems.  Firstly, a generalization of the ASEP that
allows particles also to hop to the left at a rate $q$ has been solved
\cite{PASEP}.  The phase behavior when $q<1$ is very similar to that
of the ASEP, and the rate $q$ enters into the expressions for the
current in a simple way.  The result of this would be for the pattern
of zeros in the $\alpha$-plane to shrink towards the origin as $q \to
1$.  At this point, there is a transition to a regime where the
current ceases to flow in the thermodynamic limit.  It would be of
interest to see whether the zeros of the normalization in the
complex-$q$ plane accumulate at $q=1$ since this would provide some
evidence for the generality of the Lee-Yang picture when applied to
complex transition rates.  Further evidence is provided by a study of
the percolation probability on finite directed percolation lattices
\cite{DDH} in which a similar accumulation phenomenon to that reported
here was observed.
Also, one might learn something of the
approach to the thermodynamic limit by considering how the
distribution of zeros varies
as the system size  is increased. 


\begin{acknowledgments}

We thank B.\ Derrida for insightful discussion.  R.A.B.\ thanks the
EPSRC for financial support under grant GR/R53197.

\end{acknowledgments}

\vspace{-3.75ex}



\end{document}